\documentclass{article}
\usepackage{amssymb,amsmath,graphicx,epsfig,subfig,color,hyperref,url}
\usepackage[margin=1in]{geometry}
\usepackage{cite}
\usepackage{caption}
\usepackage{mathptmx}
\usepackage{wrapfig}
\usepackage{epsfig}
\usepackage{footnote}

\def\beq {\begin{equation}}
\def\eeq {\end{equation}}
\def\bea {\begin{eqnarray}}
\def\eea {\end{eqnarray}}

\newcommand{\mzero}{m_{0}}
\newcommand{\mhalf}{m_{1/2}}
\newcommand{\azero}{A_{0}}

\def \PMET{\rm p{\!\!\!/}_T}

\newcommand{\br}{\begin{eqnarray}}
\newcommand{\er}{\end{eqnarray}}
\newcommand{\be}{\begin{equation}}
\newcommand{\ee}{\end{equation}}

\newcommand{\nut}{\widetilde\chi^0}
\newcommand{\stau}{\widetilde\tau}

\newcommand{\gsim}{\raisebox{-0.13cm}{~\shortstack{$>$ \\[-0.07cm] $\sim$}}~}

\def \PMET{p{\!\!\!/}_T}
\def\invfb{fb^{-1}}

\usepackage{color}
\usepackage{hyperref}
\usepackage{url} 
\hypersetup{linktocpage=true}
\usepackage[all]{hypcap}
\hypersetup{
   bookmarks=true,         
   unicode=false,          
   pdftoolbar=true,        
   pdfmenubar=true,        
   pdffitwindow=false,     
   pdfstartview={FitH},    
   pdftitle={My title},    
   pdfauthor={Author},     
   pdfsubject={Subject},   
   pdfcreator={Creator},   
   pdfproducer={Producer}, 
   pdfkeywords={keyword1} {key2} {key3}, 
   pdfnewwindow=true,      
   colorlinks=true,       
   linkcolor=blue,          
   citecolor=magenta,        
   filecolor=magenta,      
   urlcolor=cyan           
}
\usepackage{fancyhdr}
\begin{document}
\begin{center}
{\LARGE\bf
 Testing SUSY models for the muon g-2 anomaly via Chargino-Neutralino Pair Production at the LHC } \\[5mm]
\bigskip
{\large\sf Siba Prasad Das} $^{a},$
{\large\sf Monoranjan Guchait} $^{b}$ 
and
{\large\sf D. P. Roy} $^{c}$ 
\\ [4mm]
\bigskip
{\noindent $^{a)}$ 
Institute of Physics, Sachivalaya Marg, \\
\hspace*{0.1in} Bhubaneswar 751 005, Orissa, India.}  \\
\medskip

{\noindent $^{b)}$ 
Department of High Energy Physics, 
Tata Institute of Fundamental Research,  \\
\hspace*{0.1in} 1, Homi Bhabha Road, Mumbai 400 005, India.  }

{\noindent $^{c)}$ 
Homi Bhabha's Centre for Science Education, 
Tata Institute of Fundamental Research  \\
\hspace*{0.1in} V. N. Purav Marg, Mumbai-400088, India.  }

\end{center}

\begin{large}
\begin{abstract} 
Non-universal gaugino mass models can naturally account for the
dark matter relic density via the bulk annihilation process with relatively
light bino LSP and right sleptons in the mass range of $\sim$100 GeV, while
accommodating the observed Higgs boson mass of $\sim$125 GeV with TeV scale
squark/gluino masses. A class of these models can also account for the
observed muon g-2 anomaly via SUSY loops with wino and left sleptons in
the mass range of 400 -- 700 GeV. These models can be tested at LHC via
electroweak production of charged and neutral wino pair, leading to
robust trilepton and same sign dilepton signals. We investigate these
signals along with the standard model background for both 8 and 13 TeV
LHC runs.
\end{abstract}
\end{large}
\begin{large}
\newpage
\section{Introduction}
\label{intro}
A large part of the SUSY phenomenology over the past years
has been based on the minimal supergravity or the so-called 
constrained minimal supersymmetric standard model (CMSSM), 
which assumes universal gaugino and scalar masses $\mhalf$
and $\mzero$ at the GUT scale\cite{Kane:1998hz,Drees:2004jm,Baer:2006rs} . 
In this case the lightest superparticle (LSP), i.e., the 
dark matter, is dominantly a bino over most of the model 
parameter space. Since the bino does not carry any gauge
charge, its main annihilation process is via sfermion exchange into 
a pair of fermions. And the cosmologically compatible dark matter 
relic density requires rather small bino and sfermion masses 
$\sim$ 100 GeV. This is the so-called bulk annihilation 
region. Unfortunately, the LEP limit on the light Higgs boson 
mass, $m_h > $ 114 GeV practically rules out the bulk annihilation 
region of the CMSSM as it requires TeV scale squark/gluino masses
\cite{Beringer:1900zz}. This is further reinforced now with the 
reported discovery of Higgs boson at LHC by the ATLAS and CMS 
experiments\cite{Aad:2012tfa,Chatrchyan:2012ufa} at  
\br
m_h \simeq 125 ~{\rm GeV}.
\label{eq1}
\er  
It was shown in \cite{King:2007vh}
that the natural explanation of the cosmologically 
compatible dark matter relic density in the bulk annihilation region can be 
reconciled with the Higgs boson mass limit from LEP in a class of simple 
and well motivated MSSM with non-universal gaugino masses(NUGM) based on 
SU(5) GUT \cite{Ellis:1985jn,Drees:1985bx}.
In these models one can have relatively small bino and right slepton masses 
in the range of $\sim$ 100 GeV to account for the former along with TeV scale
squark/gluino masses to account for the latter. Moreover, these models 
can raise the Higgs boson mass to the observed range of $\sim$ 125 GeV 
with the help of a TeV scale tri-linear coupling term 
$\azero$~\cite{Mohanty:2012ri }.
More recently it was shown in \cite{Mohanty:2013soa} that some of these models 
have relatively modest wino and left slepton masses in the range of 
400-700 GeV, so that they can also account for the reported muon 
g-2 anomaly \cite{Bennett:2008dy,Davier:2010nc} via wino-left 
slepton loops\cite{Moroi:1995yh,Endo:2013bba,Gnendiger:2013pva,Fargnoli:2013zda,Fargnoli:2013zia}.

In this work we investigate the prospect of probing the above 
mentioned mass range of 400-700 GeV for wino and left sleptons 
in these models at the 8 TeV and the forthcoming 13 TeV runs 
of the LHC. Section 2 gives a brief description of the 
non-universal gaugino mass models based on SU(5) GUT. Section 3 
discusses the weak scale SUSY spectra and muon g-2 prediction 
for two such models, where the wino and left slepton 
lie in the mass range of 400-700 GeV. In particular we shall list them 
for a few benchmark points of these models for computing the LHC signal.
Section 4 describes the electroweak production of charged and neutral 
wino pair, leading to distinctive trilepton and same sign di-lepton signals 
at the LHC. It also discusses the selection cuts used in this analysis 
to extract these signals from the main standard model background. 
Section 5 discusses the results of our analysis of these two channels for 
both the 8 TeV
and the forthcoming 13 TeV runs of LHC. 
We conclude with a summary in section 6.

\section{Non-universal Gaugino Mass model in SU(5) GUT}
\label{nugm}
The gauge kinetic function responsible for the gaugino masses in the 
GUT scale Lagrangian
originates from the vacuum expectation values of the F-term of a chiral 
superfield
$\Omega$ responsible for SUSY breaking,
\br
{<{\rm F}_{\Omega}> \over {\rm M}_{planck}} {\rm \lambda_i \lambda_j}
\label{eq2}
\er  
where $\lambda_{1,2,3}$ are the U(1), SU(2) and SU(3) gaugino 
fields-- bino, wino
and gluino respectively. Since gauginos belong to the adjoint representation
 of the GUT group,
$\Omega$ and $F_{\Omega}$ can belong to any of the irreducible representations 
occurring in their symmetric product, i.e.,
\br
(24 \times 24)_{\rm sym} = 1+ 24 + 75 + 200
\label{eq3}
\er  
for the simplest GUT group SU(5). Thus for a given representation of the 
SUSY breaking superfield, the GUT scale gaugino masses are given in terms
of one mass parameter as \cite{Ellis:1985jn,Drees:1985bx}
\br 
{\rm M^G_{1,2,3} = C^n_{1,2,3} m^n_{1/2}}
\label{eq:eq4}
\er 
where 
\begin{equation}
C^1_{1,2,3}=(1,1,1), C^{24}_{1,2,3}=(-1,-3,2), C^{75}_{1,2,3}=(-5,3,1), 
C^{200}_{1,2,3}=(10,2,1).
\label{eq:eq5}
\end{equation}  
The CMSSM assumes $\Omega$ to be a singlet, leading to universal gaugino 
masses at the GUT scale. On the other hand, any of the non-singlet 
representations of $\Omega$ would imply non-universal gaugino masses 
at the GUT scale via eqs.~\ref{eq:eq4} and \ref{eq:eq5}. These non-universal
gaugino masses are known to be consistent with the universality of gauge 
couplings at the 
GUT scale \cite{Ellis:1985jn,Drees:1985bx,Chattopadhyay:2001mj}
with $\alpha_G \simeq {1 \over 25}$. The phenomenology of non-universal
gauginos arising from each of these non-singlet $\Omega$ have 
been 
widely studied~\cite{Anderson:1999uia,Huitu:1999vx,Chattopadhyay:2003yk}.

It was assumed in \cite{King:2007vh} that SUSY is broken by a 
combination of a singlet and a non-singlet superfields belonging to the 
1+24, 1+75 or 1+200 representations of SU(5). Then the GUT scale gaugino 
masses are given in terms of two mass parameters,
\begin{equation}
{\rm M^G_{1,2,3} = C^1_{1,2,3} m^1_{1/2} + C^{\ell}_{1,2,3} m^{\ell}_{1/2}, 
~~~\ell=24,75,200}. 
\label{eq:eq6}
\end{equation}  
which are determined by the two independent VEVs of the F terms of the
singlet and the non-singlet superfields.
The corresponding weak scale superparticle and Higgs boson masses are 
fixed in terms of these two gaugino mass parameters and the universal 
scalar mass parameter $\mzero$ along with the tri-linear coupling  
$\azero$ via the RGE. In these models one could access the bulk 
annihilation region of dark matter relic density, while keeping the 
Higgs boson mass above the LEP limit of 
114 GeV\cite{King:2007vh} and raise it further to the LHC value 
of $\sim$ 125 GeV with the help of a 
TeV scale  $\azero$ parameter\cite{Mohanty:2012ri}. To understand 
this result, one can equivalently consider the two independent 
gaugino mass parameter 
of eq.\ref{eq:eq6} in any of these three models to be 
$M^{G}_1$ and $M^{G}_3$. The 
corresponding weak scale gaugino masses are given to a good approximation 
by the one loop RGE, 
\begin{equation}
{\rm M_{1,2,3} = {\alpha_{1,2,3} \over \alpha_G} M^G_{1,2,3} 
\simeq {25 \over {60,30,9} }  M^G_{1,2,3}}.
\label{eq7}
\end{equation}  
Thus one can choose a relatively small $M^{G}_1$ $\sim$ 200 GeV along 
with a small $\mzero \sim$ 100 GeV to 
ensure a small weak scale bino mass $M_1$ $\sim$ 80 GeV
along with right slepton masses $\sim$ 100 GeV. Then the 
annihilation of the bino LSP pair via right slepton exchange 

\begin{equation}
\chi\chi \stackrel{\tilde\ell_R}\rightarrow  \bar \ell \ell
\label{eq:eq8}
\end{equation}  
gives the desired dark matter relic density. The other mass parameter
$M^{G}_3$ can then be raised to an appropriate level to give TeV scale
squark/gluino masses as required by the Higgs boson mass of  $\sim$ 125 GeV
and the negative squark/gluino search results from LHC.

The issue of naturalness for these nonuniversal gaugino mass models has 
been discussed in ~\cite{King:2007vh,King:2006tf} via the fine-tuning
parameters, $\Delta^\Omega$
and $\Delta^{\mathrm EW}$, required for achieving the right dark matter 
relic density and radiative EW symmetry breaking. It was found that
$\Delta^\Omega \sim$1 over the bulk annihilation region of these models,
which means there is no fine-tuning required in achieving the right dark
 matter relic density. In contrast the allowed dark matter compatible
 regions of the CMSSM or the nonuniversal scalar mass models had 1-2
 orders of magnitude higher values of this fine -tuning measure. Of course,
one has to pay the usual fine-tuning price for radiative symmetry breaking,
$\Delta^{\mathrm EW} \sim 10^2$, in the dark matter compatible regions of
all these MSSM. However, a quantitative evaluation of this fine-tuning
parameter in~\cite{King:2006tf} showed that the bulk annihilation region of the nonuniversal
gaugino mass models had one of the lowest $\Delta^{\mathrm EW}$ of them all.
Thus the low value of $\Delta^\Omega$ is achieved here without any additional
cost to the $\Delta^{\mathrm EW}$. 

Note that with given $M^{G}_1$ and $M^{G}_3$ inputs, each of the above 
three models makes a definitive prediction for $M^{G}_2$. It follows 
from eqs.~\ref{eq:eq5} and \ref{eq:eq6} that the (1+200) model predicts 
the smallest $M^{G}_2$ and hence the smallest weak scale wino and left slepton
masses among all the three models.
Hence it offers the best chance to account for a significant SUSY contribution 
to the muon g-2 anomaly, as discussed in \cite{Mohanty:2013soa}. As 
further discussed in~\cite{Mohanty:2013soa}, one can extend the analysis 
to a general non-universal gaugino mass model with three independent 
gaugino mass parameters, $M^{G}_1$, $M^{G}_2$ and $M^{G}_3$. This can 
be realized in a scenario of SUSY breaking by three superfields, 
belonging to different representations of the GUT group, 
e.g., a (1+75+200) model. 
The three gaugino mass parameters are linearly related to the VEVs 
of the F terms of these three superfields.
In this case one can have very modest wino and left slepton masses 
$\sim$ 400 GeV, so as to give a SUSY contribution to the muon g-2 
anomaly very close to its experimental central value. In the next section
we shall focus on some benchmark points from these two models, which 
can account for the muon g-2 anomaly within 2$\sigma$ level.

\section{The weak scale SUSY spectra and muon g-2 contributions in the (1+200) 
model and the general non-universal gaugino mass Model}

We have used the two-loop RGE code in SusPect \cite{Djouadi:2002ze} to 
generate the weak scale
SUSY and Higgs spectra. One should note that the $\overline{\rm MS}$ renormalization 
scheme used in the SusPect RGE code is known to predict a lower Higgs boson 
mass than the on-shell renormalization scheme used in  
FeynHiggs\cite{Arbey:2012dq} 
by 2-3 GeV~\cite{Heinemeyer:2011aa,Allanach:2004rh,Degrassi:2002fi,Harlander:2008ju,Martin:2007pg}. 
Therefore a predicted Higgs boson 
mass $\gsim$ 122 GeV
in the following tables is compatible with the reported mass 
of $\sim$ 125 GeV \cite{Aad:2012tfa,Chatrchyan:2012ufa}. 

The resulting dark matter relic density and the muon anomalous magnetic 
moment (g-2) were computed using the microOMEGAs 
code~\cite{Belanger:2010gh,Belanger:2004yn,Belanger:2001fz}. In view of 
the high precision of the dark matter relic density data
~\cite{2009ApJS..180..330K} we have considered solutions lying 
within 3$\sigma$ of its central value as in~\cite{Mohanty:2013soa}, i.e.,
\br
0.102 < \Omega h^2 < 0.123. 
\label{eq:eq9}
\er   

On the other hand the measured value  \cite{Bennett:2008dy}  of the 
muon anomalous magnetic moment excess has a relatively large uncertainty,
\begin{equation}
\Delta a_{\mu}=  (28.7 \pm 8.0) \times 10^{-10}
\label{eq:eq10}
\end{equation}  
where 
\begin{equation}
a_{\mu}= \frac{(g-2)_{\mu}}{2}.
\label{eq:eq11}
\end{equation}  
Therefore we have considered SUSY solutions 
to $a_{\mu}$~\cite{Mohanty:2013soa} lying within 2$\sigma$ of 
the central value, i.e.,
\begin{equation}
\delta a_{\mu}= \Delta a_{\mu} - a^{SUSY}_{\mu} < 2\sigma
\label{eq:eq13}
\end{equation}  
so that  $a^{SUSY}_{\mu}$ is at least of the same order as the central 
value of the experimental excess of eq.~\ref{eq:eq10}.

Explicit formulae for $a^{SUSY}_{\mu}$, arising from wino-left slepton and 
bino-right slepton loops, can be found e.g., in 
\cite{Moroi:1995yh,Endo:2013bba}. It increases 
linearly with $\tan\beta$ at constant SUSY masses. However, one has to choose 
a higher $\mzero$ at larger $tan\beta$ to maintain the $\stau_1$ mass 
in the desired range for the dark matter relic density of eq.~\ref{eq:eq9}.
The resulting increase in slepton masses compensate the linear rise 
of $a^{SUSY}_{\mu}$ with $\tan\beta$. This results in a broad peak
of $a^{SUSY}_{\mu}$ at $\tan\beta \simeq $15, which remains
nearly constant over the moderate $\tan\beta$(=10-20) 
regions~\cite{Mohanty:2013soa}.
Therefore we have chosen the following benchmark points 
at $\tan\beta$=15.
\begin{table}
\begin{center}
\begin{tabular}{|c|c|c|c|c|c|c|c|c|}
\hline
Parameters & $m_0$ & $\tan\beta$ & $A_{t0}=A_{b0}$ & $M_1^G$ & $M_2^G$ & $M_3^G$
& $a_\mu^{\rm SUSY}$ & $\delta a_\mu$\\
\hline
BP1 & 103 & 15 & -2.4 & 200 & 734.3 & 800 
& 1.59$\times 10^{-9}$ & 1.61$\sigma$ \\
BP2 & 103 & 15 & -2.4 & 200 & 822.2 & 900 
& 1.37$\times 10^{-9}$ & 1.89$\sigma$ \\
BP3 & 138 & 15 &-2.4 & 200 & 575 & 1200 
& 2.67$\times 10^{-9}$ & 0.26$\sigma$  \\
\hline
\end{tabular}
\caption{
Benchmark points of SUSY parameter space taken from Ref.~\cite{Mohanty:2013soa} 
to simulate signal process(all masses are in GeV and 
A parameters are in TeV). 
The corresponding SUSY contributions to muon anomalous magentic 
moment are shown along with their differences from 
the measured central value of eq.~\ref{eq:eq10}.}  
\label{tab:BP}
\end{center}
\end{table}

Table 1 lists three benchmark points from \cite{Mohanty:2013soa}, of which the 
BP1 and BP2 belong to the (1+200) model and BP3 
to the general non-universal gaugino mass model. The corresponding weak 
scale SUSY and Higgs spectra are listed in Table 2. 
We have checked that the rather low $\tilde \tau_1$ masses are still above 
the direct and indirect LEP limits~\cite{Beringer:1900zz}. 
The $M^{G}_3$  inputs for the BP1 and BP2 were chosen rather high to ensure 
that the resulting squark/gluino masses are well above the 8 TeV 
LHC search limits\cite{Sekmen:2014vsa}. This results in a fairly 
high values of $M^{G}_2$, so that the corresponding wino 
($\chi_{1}^{\pm}$, $\chi_{2}^{0}$) masses are $\gsim$ 600 GeV. The resulting 
$\delta a_{\mu}$ is in the range of 1.6-1.9 $\sigma$. For BP3, all the three 
gaugino mass inputs are independent. Thus we have chosen a large 
enough $M^{G}_3$
to correspond to squark/gluino masses even beyond the reach of 13 TeV 
LHC along with 
a modest $M^{G}_2$ to correspond to wino ($\chi_{1}^{\pm}$, $\chi_{2}^{0}$) 
mass $\approx$ 460 GeV. The resulting  $a^{SUSY}_{\mu}$ is 
within 0.26$\sigma$ of 
the experimental value. However, we shall see below that this benchmark point 
can be easily tested with the available 8 TeV data

\begin{table}
\begin{center}
\begin{tabular}{|c|c|c|c|c|c|c|c|c|c|c|c|c|}
\hline
Point& $\tilde{g}$ & $\tilde{q}_L$ &$\tilde q_R$  & ${\tilde t}_{1,2}$ &
$\tilde{b}_1$
& ${\tilde\ell}_{L,R}$ & ${\tilde\tau}_{1,2}$ &
${\chi}_{1}^{0}$&${\chi}_{2}^{0}$
& ${\chi}_{1}^{+} $& ${\chi}_{2}^{+}$ & $h$ \\
\hline
BP1 & 1764 & 1600 & 1540 & 820,1531 & 1311 & 479,133 & 94,480 & 80 &  593 & 593 & 1494
& 124 \\
BP2 & 1967 & 1778 & 1711 & 1012,1524 & 1490 & 531,132 & 90,532 & 80 & 666
& 666 & 1584 &124\\
BP3 & 2578 & 2252 & 2235 & 1596,1994 & 1967 & 380,159 & 96,396 & 78 & 461
& 461 & 1871 &123\\
\hline
\end{tabular}
\caption{Masses of SUSY particles(in GeV) calculated using 
SuSpect v2.41~\cite{Djouadi:2002ze} for the three benchmark points 
given in Table~\ref{tab:BP}.
}
\end{center}
\label{tab:spectrum}
\end{table}
We note from Table 2 that the left slepton ($\tilde l_L$), representing 
left selectron and smuon is always $\approx$ 20\% lighter than the 
charged and neutral wino ($\chi_{1}^{\pm}$, $\chi_{2}^{0}$).
It is a robust feature of these non-universal gaugino mass models, 
following from a small and universal $\mzero$ -- the smallness being 
required by the bulk annihilation region of dark matter relic density.
It ensures that the produced $\chi_{1}^{\pm}$, $\chi_{2}^{0}$ pair dominantly 
decay via the left sleptons, resulting in viable trilepton and same 
sign di-lepton
signals with two hard leptons. The latter signal is unaffected even if the 
left sleptons mass 
becomes very close to the wino mass. These multilepton signals will become 
unviable only if the left slepton becomes heavier than the wino, which 
will require a large $\mzero$. This means one has to sacrifice either the bulk 
annihilation region of dark matter relic density or the common scalar mass 
for the left and right sleptons. With these two resonable constraints, the 
muon g-2 satisfying SUSY models predict trilepton and even more robust 
same sign dilepton siganls at LHC.
Note that for simplicity we have assumed the same low $m_0$ value of 
the slepton sector for the squark sector as well. However, this assumption
 has no impact on our result. Assuming a large $m_0$ for the squark sector 
instead will only increase the squark masses, which has no effect however 
on the electroweak SUSY signal of our interest. 
LHC searches for other muon g-2 satisfying SUSY models have been 
discussed in \cite{Fowlie:2013oua,Freitas:2014pua,Chakraborti:2014gea}.
\section{Signal and Background}
\label{sigbg}
As discussed in the previous section the NUGM models provide a 
framework which can accommodate the bulk annihilation region of the right 
DM relic density as well as the required muon g-2. It implies that the 
wino masses are in the 
range of 400-700 GeV, which implies pair production 
of $\tilde\chi_1^\pm\nut_2$ 
at the LHC with a sizable cross section. Once this pair is 
produced in proton-proton collision, their 
cascade decays lead to the final state containing 
hard leptons along with lightest neutralinos($\nut_1$), 
which is assumed to be the 
lightest SUSY particle(LSP). 
The presence of LSPs in the final state
results in an imbalance in the measured  transverse momentum($\PMET$) 
due to its very weak 
interaction with the detector. Hence the decay channel,
\br
\tilde\chi_1^\pm \nut_2 \to (\ell^\pm \nu\nut_1)(\ell^+\ell^-\nut_1) 
\er 
with $\ell$ =e,$\mu$,
leads to a trilepton signal having same flavor opposite sign(SFOS) leptons 
or a same sign dilepton(SSDL) signal, each with   
a reasonable amount of $\PMET$. It is to be noticed that this signal
is hadronically quiet which can be exploited to get rid of 
standard model(SM) backgrounds. The dominant SM 
background is due to the WZ production with the leptonic decays 
of W and Z boson providing identical final states like the signal events.
In addition, the pair production of top quarks with the semi leptonic decays,
$t\to b W \to b\ell\nu$, and semileptonic decay of one of 
the b-quark leads to three lepton final states. Besides  
these two dominant SM backgrounds, there are other sources of 
backgrounds, e.g. from WW, WZ and W$\gamma$/Z$\gamma$ production, where 
the decay hadronic jets from W/Z can fake as leptons.
However, these  
backgrounds are expected to be very small. In the present analysis, 
we consider only the SM backgrounds due to the top pair and WZ production.
It is to be noted that in comparison with the SM backgrounds 
the leptons and $\PMET$ 
in the signal are expected to be harder, since they originate from 
comparatively more massive particles like $\tilde\chi_1^\pm$ and $\nut_2$. 
We have exploited all these signal characteristics to isolate signal 
events from 
large background samples.     
    
The signal and background events are simulated using 
{\tt PYTHIA6}~\cite{Sjostrand:2006za}. In cross section calculation we 
use CTEQ6L1~\cite{Pumplin:2002vw}
for parton distribution function setting both factorization and
renormalization scales to $\hat s$ -- the center of mass energy in the partonic
frame. In our simulation we adopt the following strategy to select events:
\\
$\bullet$ Lepton selection: As already mentioned, the signal
events are expected to contain two hard leptons due to the large mass gap 
between the left slepton and the LSP. 
So we apply hard cuts on the first two leptons
and a soft cut on the third lepton. Here leptons are arranged in  
decreasing order of $p_T$. 
For three lepton case with same flavor opposite sign(SFOS), 
\br
p_{T}^{\ell_{1,2,3}} \ge 80,50,10 {\rm~GeV}; ~|\eta^{\ell_{1,2,3}}| \le 2.5
\label{eq:l1cut}
\er
and for same sign dilepton events(SSDL), 
\br 
p_{T}^{\ell_{1,2}} \ge 50,50 {\rm~GeV}; ~|\eta^{\ell_{1,2}}| \le 2.5.
\label{eq:l2cut}
\er
The isolation of lepton is ensured by the total accompanying transverse
energy cut $E_{T}^{ac}\le 20 \%$ of the $p_T$ of the corresponding lepton,
where $E_{T}^{ac}$ is the scalar sum of transverse energies of jets
within a cone of size $\Delta R(l,j) \le 0.2 $ between jet and lepton.
These selection of cuts are very useful in suppressing the background 
events, which will be discussed later.
\\
$\bullet$ Jet selection: Jets are reconstructed using 
FastJet~\cite{Cacciari:2011ma} with jet size parameter R=0.5 and anti 
$k_T$ algorithm~\cite{Cacciari:2008gp}. Jets are selected with following 
thresholds,
\br
p_T^j \ge 30{\rm ~GeV};~|\eta^j|\le 3.0.
\label{eq:jcut}
\er
As mentioned before, signal events are hadronically quiet at the parton level
where as $t \bar t$ background events has reasonable hadronic activities. Hence,
vetoing out events having at least one jet drastically reduce the 
$t \bar t$
background by 
enormous amount, which can be observed from Tables 3-6 below.
\\
$\bullet$ In case of SFOS, we require opposite sign and same flavor
dilepton invariant mass should not lie within the range 70 - 110 GeV,
i.e if $70 < m_{ll} < 110$, events are rejected. This cut is applied 
with a goal to suppress background from WZ, where this 
dilepton invariant mass is expected to have a peak around $M_Z$.
\\
$\bullet$ The transverse missing momentum($\PMET$) is calculated adding the 
momentum of visible particles vectorially and then reverse its sign.
and it is required to be $\PMET>$150 GeV.
\\
$\bullet$ 
Another important observable, the transverse mass is found to be  
very useful to eliminate SM backgrounds without costing signal events 
too much. 
The  transverse mass is defined to be,
\br
m_T(\ell, \PMET) =\sqrt{2 p_T^{\ell} \PMET(1 - \cos\phi(p_T^{\ell},\PMET))}
\er
where $\phi(p_T^{\ell},\PMET)$ is the azimuthal angle between lepton 
and missing transverse momentum. 
In SFOS case, after applying $m_{\ell\ell}$ cut, the remaining
third lepton is used to construct $m_T(\ell_3,\PMET)$. 
The $m_T$ distribution for leptons coming from W decay either 
in top pair production or from WZ channel is expected to show a jacobian 
peak around
$M_W$; hence a cut on $m_T(\ell_3, \PMET)>$150~GeV effectively 
suppresses these backgrounds. The main suppression of the WZ background 
comes of course from the $m_{\ell\ell}$ cut.
\\
$\bullet$
For SSDL case, 
transverse mass for each lepton,
$m_T(\ell_1^\pm, \PMET)$ and $m_T(\ell_2^\pm, \PMET)$ are constructed
and selection cuts are applied separately requiring these to be more than
100 GeV. These cuts are very useful in suppressing the $t\bar t $ and 
WZ backgrounds. In this case a transverse mass cut of the dilepton system
with the $\PMET$, $m_T(\ell_1+\ell_2,\PMET) >$ 125 GeV, also helps 
to suppress these backgrounds further.    

The signal and backgrounds are simulated for both LHC energies 8 TeV 
and as well as 13 TeV which is expected to be the Run 2 LHC energy 
in the next year.
For illustration, the signal rates are estimated for the three representative 
choices of parameter space as shown in Table~\ref{tab:BP}, and 
the corresponding mass spectra as presented in Table~2. 
\section{Results and Discussions}
\label{result}
We present the summary of events in Table 3-6 
for both 8 and 13 TeV energies simulating both signal and SM 
backgrounds $t \bar t$ and WZ, adopting the strategy as 
described in the previous section.
\begin{table}
\begin{center}
\begin{tabular}{|c|c|c|c|c|c|c|c||c|c|}
\hline
Proc & NoE & $\sigma$(fb) & 3${\ell}$ & $\PMET$&$m_{ll} \neq$
& $m_T(\ell_3,\PMET)$ & Jet Veto(JV) &
\multicolumn{2}{|c|}{$\sigma \epsilon_{ac}$(fb)}\\
& & &  & $\ge$150 & $M_Z \pm 20 $ & $\ge150$ &  & No JV & JV \\
\hline
BP1 & 50K & 1.8 & 9354 & 7393 & 6728 & 5811 & 3407 & 0.20 & 0.12 \\
BP2&50K& 0.88& 9591 &   7909 & 7318 & 6409  & 3707 & 0.11 & 0.06 \\
BP3&50K & 11.3 &7254  &  4996  & 4263 &  3543 & 2180 & 0.80 & 0.49 \\
\hline
$t \bar t$:0-200  &2M   & 88400&     689& 15&  13& 1& 0& 0.04 & 0.00\\
$t \bar t$:200-500&0.2M & 9710 & 238 & 22 & 18 & 7 & 0 & 0.33 & 0.00 \\
$t \bar t$:500-up &$ 10^5$ &130  & 288 & 128 &119 & 44 & 0 & 0.05 & 0.0\\
\hline
$W^{\pm}Z$& 0.45M & 13000 & 525 & 32 & 10 & 10 & 6 &0.28 &0.17    \\
\hline
\end{tabular}
\caption{
Event summary for trilepton final state with same flavor opposite sign
(SFOS) corresponding to signal and SM backgrounds at 8 TeV energy along
with production cross section in LO(third column).
The last two columns show the normalized cross-section
($\sigma \times$ acceptance efficiency($\epsilon_{ac}$)) without and with
jet veto(JV). The $t \bar t$ is simulated for three $\hat p_T$ bins.}
\end{center}
\end{table}
Table 3 presents the number of trilepton events for 8 TeV energy 
after each set of cuts as shown on top of each columns.  
The 2nd and 3rd columns show the 
number of events(NoE) simulated and leading order(LO) cross 
sections(in fb) respectively for each process, where as the fourth 
column presents the 
number of events having 3 leptons in the final states 
passing cut, eq.~\ref{eq:l1cut}. 
Note that the $t\bar t$ events are simulated for three $\hat p_T$ bins
to consider statistics appropriately in different phase space regions.
Here $\hat p_T$ is the transverse momentum of top quark pair in partonic
frame. 
Notice that selection cuts on $\PMET$ and $m_T(\ell_3,\PMET)$ are very
effective to suppress background events, in particular $t \bar t$ 
events, where as dilepton invariant mass($m_{\ell\ell}$) cut suppresses 
mainly WZ background,  with little effect on signal 
events. Eventually the jet veto(JV) criteria, i.e.  
reject events if there exist jets in the final states, reduces the $t \bar t$ 
backgrounds drastically with a mild effect on signal events.
As noted earlier, 
in signal events presence of jets are mainly due to initial state 
radiation; and hence the signal events are not expected to have 
many hard jets unlike the $t \bar t$  
background.
The last two columns display the final cross sections multiplying  
by acceptance efficiency for both cases, with and without jet veto. 
Clearly, jet veto completely brings down the top backgrounds to a 
negligible level, but residual WZ background remains.       

\begin{table}
\begin{center}
\begin{tabular}{|c|c|c|c|c|c|c|c|c|c|c|}
\hline
Proc & NoE & $\sigma$(fb) & $n_{\ell}=2$ & $\PMET$ 
&$m_T({\ell_1},\PMET)$  &$ m_T({\ell_2},\PMET)$& $m_T({\ell_1+\ell_2},\PMET)$
&$n_{j}=0$&
\multicolumn{2}{|c|}{$\sigma\times\epsilon$(fb)}\\
&&&SSDL&$\ge$150& $\ge$100 &$\ge$100& $\ge$125 &JetVeto(JV)&No JV&JV\\
\hline
BP1 & 50K & 1.8  & 9249 & 7180 & 6953 & 6082 & 5921 & 3338  & 0.21 &0.12 \\
BP2 & 50K &  0.88 & 9692 & 7876 & 7642 & 6753 & 6790 & 3678 & 0.12 &0.06 \\
BP3 & 50K &  11.3   & 6541 & 4260 & 4131 & 3548 & 3469 & 2031 &0.78 & 0.45 \\
\hline
$t \bar t$0-200 & 2M   & 88400 & 148 & 3 & 1 & 0 &0 &0 &0 &0  \\
$t \bar t$200-500&0.2M & 9710  &75  & 10 & 5 &0 &0  &0 & 0&0  \\
$t \bar t$500-up&0.1M  &  130  & 474 & 203 &56 & 2  &1  &0 &0.001 & 0  \\
\hline
$W^{\pm}Z$& .45M & 13000 &568 & 14 & 12 & 3 & 1& 1 &0.003   &0.003   \\
\hline
\end{tabular}
\caption{
Same as Table 3, but for same sign
dilepton case(SSDL).
}
\end{center}
\label{tab:ssdl8}
\end{table}
In Table 4 we show event summary for SSDL case at 8 TeV
energy. In this case we apply same set of selection cuts as discussed before, 
but in addition two more  
selection cuts $m_T(\ell_1,\PMET)$ and $m_T(\ell_2,\PMET)$ 
are used with a purpose to suppress
mainly WZ background. It is motivated by the fact that in WZ channel 
two leptons always come from W and Z decays, and the one coming from W decay
will not kinematically pass the $m_T>$100 GeV cut. In contrast  
the signal events, where leptons originate from
heavier $\chi_1^\pm$ and $\nut_2$ decays pass the $m_T>$100 GeV cut for 
both the leptons.
Finally, as indicated by the Table 4, 
the level of backgrounds cross sections turn out to be negligible. 
So the discovery limit in the SSDL channels is determined essentially by the 
signal size.

\begin{table}
\begin{center}
\begin{tabular}{|c|c|c|c|c|c|c|c||c|c|}
\hline
Proc & NoE &$\sigma$(fb) & 3${\ell}$ &$\PMET$ &$m_{ll} \neq$ &
$m_T(\ell_3,\PMET)$ &$\rm{n_j}=0$ &
\multicolumn{2}{|c|}{$\sigma\times\epsilon$(fb)}\\
&&&&$>$150&$m_Z \pm 20$& $>150$ &Jet Veto(JV)  & No JV &  JV\\
\hline
BP1 &  50K  & 7.4 & 8880 & 7123 & 6459 & 5519 & 2883 & 0.81 & 0.43 \\
BP2 &  50K & 4.2 &  9290 & 7753 & 7168 & 6228 & 3084&  0.52 &0.26 \\
BP3 &  50K & 38 & 6941 & 4882 & 4196 & 3470 & 1885 &  2.64 & 1.42\\
\hline
$t \bar t$0-200 & 30M & 362000 & 10983 & 543 & 468 &48 & 1 & 0.57 & 0.012   \\
$t \bar t$200-500 & 4M & 40000 & 4286 & 615 &549  &175  &6  &1.75 &.06   \\
$t \bar t$500-Inf & 0.1M & 810 & 300 & 129 & 116 & 46 & 0 &0.37 &0   \\
\hline
$W^{\pm}Z$ &4M& 26000 &4330  & 902 & 97 & 84 & 45 & 0.54  &0.29    \\
\hline
\end{tabular}
\caption{ Event summary for SFOS case, same as 
Table 3 but for 13 TeV energy.}
\end{center}
\label{tab:sfos13}
\end{table}
Similarly we simulate signal and background events for 13 TeV energy 
using same set of cuts for both SFOS and SSDL cases, which are presented  
in Tables 5 and 6 respectively.  
The pattern of suppression of background events
with respect to signal events are more or less the same as observed
before. However, 
effect of jet 
veto kills signal events a little more than at 8 TeV due to the fact that 
hadronic activities from ISR/FSR are more at this higher energy.   
\begin{table}
\begin{center}
\begin{tabular}{|c|c|c|c|c|c|c|c|c|c|c|}
\hline
Proc & Evt & $\sigma$(fb) & $n_{\ell}=2$ & $\PMET$
&$m_T({\ell_1},\PMET)$  &$ m_T({\ell_2},\PMET)$& $m_T({\ell_1+\ell_2},\PMET)$
&$n_{j}=0$&
\multicolumn{2}{|c|}{$\sigma\times\epsilon$(fb)}\\
&&&SSDL&$\ge$150& $\ge$100 &$\ge$100& $\ge$125 &JetVeto(JV)&No JV&JV\\
\hline
BP1 & 50K &7.4   & 8803 & 6961 & 6721 & 5882 & 5710 & 2925  &0.84 &0.44  \\
BP2 & 50K &4.2   & 9663 & 7942 & 7691 & 6772 & 6582 & 3145 &0.55  &0.26  \\
BP3 & 50K &38   & 6420 & 4295 & 4131 & 3481 & 3383 & 1821 &2.57  & 1.38 \\
\hline
$t \bar t$0-200 & 30M &362000  & 2566  & 165 & 30 & 2 & 1  & 0 & 0.01 & 0  \\
$t \bar t$200-500& 4M &40000   & 1835  & 297 & 129 & 17 &13  & 0 & 0.13& 0  \\
$t \bar t$500-up&0.1M  & 810   & 546 & 255 & 74 & 9  & 5  & 0  & 0.04  & 0  \\
\hline
$W^{\pm}Z$&4M   & 26000 & 4607 & 157 & 104 & 2 &  2 & 1 &0.013   &.006   \\
\hline
\end{tabular}
\caption{
Event summary for SSDL, same as Table 4 
but for 13 TeV energy.}
\end{center}
\label{tab:ssdl13}
\end{table}

\begin{table}
\begin{center}
\begin{tabular}{|c|c|c|c|c|}
\hline
Process & 8 TeV & 8 TeV & 13 TeV & 13 TeV\\ 
& No JV &  JV & No JV &  JV \\
\hline
$t\bar t$ & 0.67 & -- & 4.3  &0.11 \\
WZ      &   0.48   & 0.29   & 0.92 & 0.5  \\
\hline
Total Bg & 1.15 &  0.29 & 5.22 & 0.61    \\
\hline
BP1 & 0.3 & 0.18  & 1.21  & 0.64 \\
$\frac{S}{\sqrt{B}}$  &1.25 &1.5 &5.3  & 8.2\\
\hline
BP2 & 0.165 &  0.09 &0.78  &0.39 \\
$\frac{S}{\sqrt{B}}$ &0.68  &0.74  & 3.4 & 5 \\
\hline
BP3 &  1.2 & 0.73 & 3.96 & 2.13 \\
$\frac{S}{\sqrt{B}}$  &5.0  & 6. &17.3  &27  \\
\hline
\end{tabular}
\caption{Total background and signal cross sections(in fb) for 
trilepton final states after all selection cuts corresponding to 
each benchmark point.  
Note that these cross-sections are obtained by multiplying the background and
signal cross-sections of Tables 3 and 5 by the appropriate K-factors
as described in the text.
The significance are computed for 
integrated luminosities 20$\invfb$ and 100$\invfb$ for 8 and 13 TeV
energies respectively. }
\end{center}
\label{tab:sfos}
\end{table}

\begin{table}
\begin{center}
\begin{tabular}{|c|c|c|c|c|}
\hline
Process & 8 TeV & 8 TeV & 13 TeV & 13 TeV\\
& No JV &  JV & No JV &  JV \\
\hline
$t\bar t$ &.002  &0 &0.28 &0 \\
WZ      &  0.005 &0.005  &0.022 & 0.01 \\
\hline
Total Bg &0.007 &.005&0.30 & 0.01\\
\hline
BP1 & 0.31 &0.18   &1.26   &0.66  \\
S  &6 &3.6 &126  &66. \\
\hline
BP2 &0.18  &0.09   &0.82  &0.39 \\
S &3.6 &1.8  &82  &39  \\
\hline
BP3 &1.17  &0.67  &3.85  &2.07  \\
S & 23 &13  & 385 & 207 \\
\hline
\end{tabular}
\caption{Same as Table 7, but for SSDL case.
In this case we show the expected number of signal events for the integrated
luminosities of 20$\invfb$ and 100$\invfb$ for 8 and 13 TeV respectively.
The corresponding $S/\sqrt{B} \ge$10 for all cases.
}
\end{center}
\label{tab:ssdl}
\end{table}
Finally we summarize our results presenting signal and background cross 
sections along with the signal significance($S/\sqrt{B}$)  
in Tables 7 and 8 for SFOS and SSDL cases 
respectively. The significance are estimated for integrated luminosity
20$\invfb$ and 100$\invfb$ corresponding to 8 and 13 TeV LHC energies.
In calculating both signal and background cross sections, we have 
taken into account the next to leading order effect by multiplying 
the K-factors for each cases.
For example, for $t\bar t$ and WZ processes, we multiply  
cross sections by 1.6~\cite{Kidonakis:2011tg} and 
1.7~\cite{Campbell:2011bn}, where as for signal  
it is 1.5~\cite{Beenakker:1999xh}. Although these K-factors are derived
for 14 TeV energy, 
they are not expected to be very different at 8 and 13 TeV.

Table 7 shows the summary of the tri-lepton(SFOS) channel results.
We see from this Table that the BP3 corresponding to modest 
wino ($\chi_{1}^{\pm}$, $\chi_{2}^{0}$) mass of $\approx$ 460 GeV
can be probed at 5(6)$\sigma$ level with 24(15) trilepton signal 
events without(with) jet veto 
from the available 20 $fb^{-1}$ data at 8 TeV. 
Even without a dedicated search with the model, it may be 
reasonable to assume that trilepton signal of this size 
could not be missed in generic search of chargino-neutralino pair production 
events. On the other hand for BP1 and BP2, corresponding to wino 
mass $m_{\chi_{1}^{\pm}}$, $m_{\chi_{2}^{0}}$ $\gsim$ 600 GeV, 
one expects only a couple of trilepton signal events 
at a significance level $< 2\sigma$ with the available 20 $fb^{-1}$ 
data at 8 TeV. But with the 100 $fb^{-1}$ data at 13 TeV one can probe
BP1(BP2) at a significance level of $\sim$ $8\sigma$($5\sigma$)  
with 60(40) trilepton signal events. This means that even with a
20 $fb^{-1}$ data at 13 TeV a negative search result can rule out wino
($\chi_{1}^{\pm}$, $\chi_{2}^{0}$)  masses upto 600-700 GeV at $> 2\sigma$ 
level. This will essentially cover the non-universal gaugino mass models 
satisfying muon g-2 anomaly up to $2\sigma$ level. 

The summary of the corresponding results for the SSDL channel 
is shown in Table.8. In this case possibility of observing signal events 
is more promising due to the presence of tiny backgrounds.
Indeed the $S/\sqrt{B}$ ratio is $\ge$10 for all the cases studied 
here so that the discovery limit is essentially determined by the 
number of signal events. Therefore, we show this number here instead of
the $S/\sqrt{B}$ ratio.
We see from this table that the BP3, corresponding 
to modest wino mass of $\approx$ 460 GeV, can be probed  
with 23(13)  
SSDL signal events without(with) jet veto from the available 
20 $fb^{-1}$  data at 8 TeV. 
Again it is reasonable to assume that a signal of this size cannot be 
missed even in a generic search of chargino-neutralino pair production 
with this data. For BP1 and BP2, corresponding to wino mass 
$\gsim$ 600 GeV, one expects only 4-6 and 2-3 SSDL signal events
respectively. This falls short of a conservative discovery limit of at least
5-6 events.
With the 100 $fb^{-1}$  data at 13 TeV
one can probe BP1(BP2) at a  significance level of  
$\approx$ $26-66\sigma$($15-39\sigma$) 
with 125-65(80-40) SSDL signal events. Thus even with a 20 $fb^{-1}$ 
data at 13 TeV a negative search results can rule out 
wino($\chi_{1}^{\pm}$, $\chi_{2}^{0}$) masses up to 600-700 GeV 
at $>$ $5\sigma$ level. Thus one can unambiguously 
probe the muon g-2 anomaly satisfying non-universal gaugino mass models 
at the 13 TeV LHC using either the trilepton or SSDL channels.
The SSDL channel has the advantage of a very small background.
Besides the SSDL channel also has the advantage of 
being viable even when the left slepton mass comes very 
close the wino mass, as discussed earlier.

Recently the ATLAS collaboration have published the analysis of 
their 20 $fb^{-1}$  data at 8 TeV for chargino pair production signal 
in the unlike sign dilepton channel~\cite{Aad:2014vma}. For a
80 GeV LSP ($\nut_1$), they show an expected exclusion region 
up to $m_{\chi_{1}^{\pm}}$=450 -550 GeV at the 95\% C.L. 
($\simeq$ $2\sigma$), 
which is similar to our BP3. There is a preliminary CMS result of search 
for electroweak chargino-neutralino pair 
production in the tri-lepton channel using their 20 $fb^{-1}$ data 
at 8 TeV \cite{Khachatryan:2014qwa}.
While most of their analysis focuses on a left slepton mass 
midway between the  $\chi_{1}^{0}$ and $\chi_{2}^{0}$(=$\chi_{1}^{\pm}$) 
masses, there is one figure( Fig.15b in Ref.~\cite{Khachatryan:2014qwa})
showing the 95\% C.L. 
exclusion regions in $\chi_{1}^{0}$ and $\chi_{2}^{0}$ masses for 
left slepton mass close to the latter. The edge of their expected 
95\% C.L. ($\simeq$ $2\sigma$) exclusion region for $m_{\chi_{1}^{0}}$=80 
GeV touches $m_{\chi_{2}^{0}}$=600 GeV which is close to our BP1. 
Thus their expected $2\sigma$ exclusion limit is stronger than our 
estimated $1.5 \sigma$ exclusion for BP1 in Table 7. The main reason 
for this seems to be their use of b-jet veto instead of a general 
jet veto, so that they can suppress the $t \bar t$ background without 
sacrificing the SUSY signal. Their b-jet veto criteria have been 
tuned to their $t \bar t$ data. Having no access to this data, 
we had to rely on the general jet veto to suppress 
the $t \bar t$ background. We hope the CMS collaboration will 
do a dedicated analysis of their 8 TeV data for chargino-neutralino 
pair production in tri-lepton and SSDL channels in these 
non-universal gaugino mass models, where the electroweak super particle 
masses are fairly well constrained by the dark matter relic density 
and the muon g-2 anomaly.

\section{Summary}
Non-universal gaugino mass models can naturally account for the dark matter 
relic density via the bulk annihilation process with relatively light bino 
LSP and right sleptons in the mass rage of $\sim$100 GeV, while accommodating 
the observed Higgs boson mass of $\sim$ 125 GeV with TeV scale squark/gluino 
masses. Some of these models can also account for the observed muon g-2 
anomaly via SUSY loops with wino and left sleptons in the mass range of 
400-700 GeV. We have investigated the prospect of testing these models via 
electroweak production of charged and neutral wino pairs at the LHC. 
The left slepton masses in these models are predicted to lie typically 
$\sim$20\% below the wino mass. Thus one expects robust trilepton and 
same sign dilepton signals of these models arising from the cascade 
decays of the charged and neutral wino pair via the left sleptons. 
In particular the SSDL signal holds even when the left slepton mass lies 
very close to the wino mass. It also has the advantage of a very small 
standard model background. Our simulation study shows that the available 
8 TeV LHC data is adequate to probe the wino mass range of 400-500 GeV in 
both the trilepton and the SSDL channels. This mass range of wino covers 
the muon g-2 range within 0-1$\sigma$ of its observed central value. Moreover 
the probe can be extended to the wino mass range of 600-700 GeV with the 
13 TeV LHC data, which covers the muon g-2 range up to 2$\sigma$ of its 
central value. Thus the non-universal gaugino mass models satisfying the 
observed dark matter relic density and the muon g-2 anomaly can be 
unambiguously tested via electroweak production of the charged and neutral 
wino pair at the forthcoming 13 TeV run of LHC.

\section{Acknowledgement}
The work of DPR was partly supported by the senior scientist fellowship of the
Indian National Science Academy. The authors are also thankful to the organisers
of the '13th Workshop on High Energy Physics Phenomenology(WHEPP 13)' held
at Puri, Odisha, 12-21st December,2013, where this project was started.
\end{large}
\bibliography{paper.bib}{}

\begin{thebibliography}{10}

\bibitem{Kane:1998hz}
G.~L. Kane,
``{Perspectives on supersymmetry},''.

\bibitem{Drees:2004jm}
M.~Drees, R.~Godbole, and P.~Roy,
``{Theory and phenomenology of sparticles: An account of four-dimensional N=1
  supersymmetry in high energy physics},''.

\bibitem{Baer:2006rs}
H.~Baer and X.~Tata,
``{Weak scale supersymmetry: From superfields to scattering events},''.

\bibitem{Beringer:1900zz}
{\bfseries Particle Data Group} Collaboration, J.~Beringer {\em et~al.},
  ``{Review of Particle Physics (RPP)},''
\href{http://dx.doi.org/10.1103/PhysRevD.86.010001}{{\em Phys.Rev.} {\bfseries
  D86} (2012) 010001}.

\bibitem{Aad:2012tfa}
{\bfseries ATLAS Collaboration} Collaboration, G.~Aad {\em et~al.},
  ``{Observation of a new particle in the search for the Standard Model Higgs
  boson with the ATLAS detector at the LHC},''
  \href{http://dx.doi.org/10.1016/j.physletb.2012.08.020}{{\em Phys.Lett.}
  {\bfseries B716} (2012) 1--29},
\href{http://arxiv.org/abs/1207.7214}{{\ttfamily arXiv:1207.7214 [hep-ex]}}.

\bibitem{Chatrchyan:2012ufa}
{\bfseries CMS Collaboration} Collaboration, S.~Chatrchyan {\em et~al.},
  ``{Observation of a new boson at a mass of 125 GeV with the CMS experiment at
  the LHC},'' \href{http://dx.doi.org/10.1016/j.physletb.2012.08.021}{{\em
  Phys.Lett.} {\bfseries B716} (2012) 30--61},
\href{http://arxiv.org/abs/1207.7235}{{\ttfamily arXiv:1207.7235 [hep-ex]}}.

\bibitem{King:2007vh}
S.~King, J.~Roberts, and D.~Roy, ``{Natural dark matter in SUSY GUTs with
  non-universal gaugino masses},''
  \href{http://dx.doi.org/10.1088/1126-6708/2007/10/106}{{\em JHEP} {\bfseries
  0710} (2007) 106},
\href{http://arxiv.org/abs/0705.4219}{{\ttfamily arXiv:0705.4219 [hep-ph]}}.

\bibitem{Ellis:1985jn}
J.~R. Ellis, K.~Enqvist, D.~V. Nanopoulos, and K.~Tamvakis, ``{Gaugino Masses
  and Grand Unification},''
\href{http://dx.doi.org/10.1016/0370-2693(85)91591-6}{{\em Phys.Lett.}
  {\bfseries B155} (1985) 381}.

\bibitem{Drees:1985bx}
M.~Drees, ``{Phenomenological Consequences of $N=1$ Supergravity Theories With
  Nonminimal Kinetic Energy Terms for Vector Superfields},''
\href{http://dx.doi.org/10.1016/0370-2693(85)90442-3}{{\em Phys.Lett.}
  {\bfseries B158} (1985) 409}.

\bibitem{Mohanty:2012ri}
S.~Mohanty, S.~Rao, and D.~Roy, ``{Predictions of a Natural SUSY Dark Matter
  Model for Direct and Indirect Detection Experiments},''
  \href{http://dx.doi.org/10.1007/JHEP11(2012)175}{{\em JHEP} {\bfseries 1211}
  (2012) 175},
\href{http://arxiv.org/abs/1208.0894}{{\ttfamily arXiv:1208.0894 [hep-ph]}}.

\bibitem{Mohanty:2013soa}
S.~Mohanty, S.~Rao, and D.~Roy, ``{Reconciling the muon $g-2$ and dark matter
  relic density with the LHC results in nonuniversal gaugino mass models},''
  \href{http://dx.doi.org/10.1007/JHEP09(2013)027}{{\em JHEP} {\bfseries 1309}
  (2013) 027},
\href{http://arxiv.org/abs/1303.5830}{{\ttfamily arXiv:1303.5830 [hep-ph]}}.

\bibitem{Bennett:2008dy}
{\bfseries Muon (g-2) Collaboration} Collaboration, G.~Bennett {\em et~al.},
  ``{An Improved Limit on the Muon Electric Dipole Moment},''
  \href{http://dx.doi.org/10.1103/PhysRevD.80.052008}{{\em Phys.Rev.}
  {\bfseries D80} (2009) 052008},
\href{http://arxiv.org/abs/0811.1207}{{\ttfamily arXiv:0811.1207 [hep-ex]}}.

\bibitem{Davier:2010nc}
M.~Davier, A.~Hoecker, B.~Malaescu, and Z.~Zhang, ``{Reevaluation of the
  Hadronic Contributions to the Muon g-2 and to alpha(MZ)},''
  \href{http://dx.doi.org/10.1140/epjc/s10052-012-1874-8,
  10.1140/epjc/s10052-010-1515-z}{{\em Eur.Phys.J.} {\bfseries C71} (2011)
  1515},
\href{http://arxiv.org/abs/1010.4180}{{\ttfamily arXiv:1010.4180 [hep-ph]}}.

\bibitem{Moroi:1995yh}
T.~Moroi, ``{The Muon anomalous magnetic dipole moment in the minimal
  supersymmetric standard model},''
  \href{http://dx.doi.org/10.1103/PhysRevD.53.6565,
  10.1103/PhysRevD.56.4424}{{\em Phys.Rev.} {\bfseries D53} (1996) 6565--6575},
\href{http://arxiv.org/abs/hep-ph/9512396}{{\ttfamily arXiv:hep-ph/9512396
  [hep-ph]}}.

\bibitem{Endo:2013bba}
M.~Endo, K.~Hamaguchi, S.~Iwamoto, and T.~Yoshinaga, ``{Muon $g-2$ vs LHC in
  Supersymmetric Models},''
  \href{http://dx.doi.org/10.1007/JHEP01(2014)123}{{\em JHEP} {\bfseries 1401}
  (2014) 123},
\href{http://arxiv.org/abs/1303.4256}{{\ttfamily arXiv:1303.4256 [hep-ph]}}.

\bibitem{Gnendiger:2013pva}
C.~Gnendiger, D.~Stöckinger, and H.~Stöckinger-Kim, ``{The electroweak
  contributions to $(g-2)_\mu$ after the Higgs boson mass measurement},''
  \href{http://dx.doi.org/10.1103/PhysRevD.88.053005}{{\em Phys.Rev.}
  {\bfseries D88} no.~5, (2013) 053005},
\href{http://arxiv.org/abs/1306.5546}{{\ttfamily arXiv:1306.5546 [hep-ph]}}.

\bibitem{Fargnoli:2013zda}
H.~Fargnoli, C.~Gnendiger, S.~Paßehr, D.~Stöckinger, and H.~Stöckinger-Kim,
  ``{Non-decoupling two-loop corrections to $(g-2)$$_{\mu}$ from
  fermion/sfermion loops in the MSSM},''
  \href{http://dx.doi.org/10.1016/j.physletb.2013.09.034}{{\em Phys.Lett.}
  {\bfseries B726} (2013) 717--724},
\href{http://arxiv.org/abs/1309.0980}{{\ttfamily arXiv:1309.0980 [hep-ph]}}.

\bibitem{Fargnoli:2013zia}
H.~Fargnoli, C.~Gnendiger, S.~Paßehr, D.~Stöckinger, and H.~Stöckinger-Kim,
  ``{Two-loop corrections to the muon magnetic moment from fermion/sfermion
  loops in the MSSM: detailed results},''
  \href{http://dx.doi.org/10.1007/JHEP02(2014)070}{{\em JHEP} {\bfseries 1402}
  (2014) 070},
\href{http://arxiv.org/abs/1311.1775}{{\ttfamily arXiv:1311.1775 [hep-ph]}}.

\bibitem{Chattopadhyay:2001mj}
U.~Chattopadhyay and P.~Nath, ``{b - tau unification, g(mu) - 2, the b ---gt; s
  + gamma constraint and nonuniversalities},''
  \href{http://dx.doi.org/10.1103/PhysRevD.65.075009}{{\em Phys.Rev.}
  {\bfseries D65} (2002) 075009},
\href{http://arxiv.org/abs/hep-ph/0110341}{{\ttfamily arXiv:hep-ph/0110341
  [hep-ph]}}.

\bibitem{Anderson:1999uia}
G.~Anderson, H.~Baer, C.-h. Chen, and X.~Tata, ``{The Reach of Fermilab
  Tevatron upgrades for SU(5) supergravity models with nonuniversal gaugino
  masses},'' \href{http://dx.doi.org/10.1103/PhysRevD.61.095005}{{\em
  Phys.Rev.} {\bfseries D61} (2000) 095005},
\href{http://arxiv.org/abs/hep-ph/9903370}{{\ttfamily arXiv:hep-ph/9903370
  [hep-ph]}}.

\bibitem{Huitu:1999vx}
K.~Huitu, Y.~Kawamura, T.~Kobayashi, and K.~Puolamaki, ``{Phenomenological
  constraints on SUSY SU(5) GUTs with nonuniversal gaugino masses},''
  \href{http://dx.doi.org/10.1103/PhysRevD.61.035001}{{\em Phys.Rev.}
  {\bfseries D61} (2000) 035001},
\href{http://arxiv.org/abs/hep-ph/9903528}{{\ttfamily arXiv:hep-ph/9903528
  [hep-ph]}}.

\bibitem{Chattopadhyay:2003yk}
U.~Chattopadhyay and D.~Roy, ``{Higgsino dark matter in a SUGRA model with
  nonuniversal gaugino masses},''
  \href{http://dx.doi.org/10.1103/PhysRevD.68.033010}{{\em Phys.Rev.}
  {\bfseries D68} (2003) 033010},
\href{http://arxiv.org/abs/hep-ph/0304108}{{\ttfamily arXiv:hep-ph/0304108
  [hep-ph]}}.

\bibitem{King:2006tf}
S.~King and J.~Roberts, ``{Natural implementation of neutralino dark matter},''
  \href{http://dx.doi.org/10.1088/1126-6708/2006/09/036}{{\em JHEP} {\bfseries
  0609} (2006) 036},
\href{http://arxiv.org/abs/hep-ph/0603095}{{\ttfamily arXiv:hep-ph/0603095
  [hep-ph]}}.

\bibitem{Djouadi:2002ze}
A.~Djouadi, J.-L. Kneur, and G.~Moultaka, ``{SuSpect: A Fortran code for the
  supersymmetric and Higgs particle spectrum in the MSSM},''
  \href{http://dx.doi.org/10.1016/j.cpc.2006.11.009}{{\em Comput.Phys.Commun.}
  {\bfseries 176} (2007) 426--455},
\href{http://arxiv.org/abs/hep-ph/0211331}{{\ttfamily arXiv:hep-ph/0211331
  [hep-ph]}}.

\bibitem{Arbey:2012dq}
A.~Arbey, M.~Battaglia, A.~Djouadi, and F.~Mahmoudi, ``{The Higgs sector of the
  phenomenological MSSM in the light of the Higgs boson discovery},''
  \href{http://dx.doi.org/10.1007/JHEP09(2012)107}{{\em JHEP} {\bfseries 1209}
  (2012) 107},
\href{http://arxiv.org/abs/1207.1348}{{\ttfamily arXiv:1207.1348 [hep-ph]}}.

\bibitem{Heinemeyer:2011aa}
S.~Heinemeyer, O.~Stal, and G.~Weiglein, ``{Interpreting the LHC Higgs Search
  Results in the MSSM},''
  \href{http://dx.doi.org/10.1016/j.physletb.2012.02.084}{{\em Phys.Lett.}
  {\bfseries B710} (2012) 201--206},
\href{http://arxiv.org/abs/1112.3026}{{\ttfamily arXiv:1112.3026 [hep-ph]}}.

\bibitem{Allanach:2004rh}
B.~Allanach, A.~Djouadi, J.~Kneur, W.~Porod, and P.~Slavich, ``{Precise
  determination of the neutral Higgs boson masses in the MSSM},''
  \href{http://dx.doi.org/10.1088/1126-6708/2004/09/044}{{\em JHEP} {\bfseries
  0409} (2004) 044},
\href{http://arxiv.org/abs/hep-ph/0406166}{{\ttfamily arXiv:hep-ph/0406166
  [hep-ph]}}.

\bibitem{Degrassi:2002fi}
G.~Degrassi, S.~Heinemeyer, W.~Hollik, P.~Slavich, and G.~Weiglein, ``{Towards
  high precision predictions for the MSSM Higgs sector},''
  \href{http://dx.doi.org/10.1140/epjc/s2003-01152-2}{{\em Eur.Phys.J.}
  {\bfseries C28} (2003) 133--143},
\href{http://arxiv.org/abs/hep-ph/0212020}{{\ttfamily arXiv:hep-ph/0212020
  [hep-ph]}}.

\bibitem{Harlander:2008ju}
R.~Harlander, P.~Kant, L.~Mihaila, and M.~Steinhauser, ``{Higgs boson mass in
  supersymmetry to three loops},''
  \href{http://dx.doi.org/10.1103/PhysRevLett.101.039901,
  10.1103/PhysRevLett.100.191602}{{\em Phys.Rev.Lett.} {\bfseries 100} (2008)
  191602},
\href{http://arxiv.org/abs/0803.0672}{{\ttfamily arXiv:0803.0672 [hep-ph]}}.

\bibitem{Martin:2007pg}
S.~P. Martin, ``{Three-loop corrections to the lightest Higgs scalar boson mass
  in supersymmetry},'' \href{http://dx.doi.org/10.1103/PhysRevD.75.055005}{{\em
  Phys.Rev.} {\bfseries D75} (2007) 055005},
\href{http://arxiv.org/abs/hep-ph/0701051}{{\ttfamily arXiv:hep-ph/0701051
  [hep-ph]}}.

\bibitem{Belanger:2010gh}
G.~Belanger, F.~Boudjema, P.~Brun, A.~Pukhov, S.~Rosier-Lees, {\em et~al.},
  ``{Indirect search for dark matter with micrOMEGAs2.4},''
  \href{http://dx.doi.org/10.1016/j.cpc.2010.11.033}{{\em Comput.Phys.Commun.}
  {\bfseries 182} (2011) 842--856},
\href{http://arxiv.org/abs/1004.1092}{{\ttfamily arXiv:1004.1092 [hep-ph]}}.

\bibitem{Belanger:2004yn}
G.~Belanger, F.~Boudjema, A.~Pukhov, and A.~Semenov, ``{micrOMEGAs: Version
  1.3},'' \href{http://dx.doi.org/10.1016/j.cpc.2005.12.005}{{\em
  Comput.Phys.Commun.} {\bfseries 174} (2006) 577--604},
\href{http://arxiv.org/abs/hep-ph/0405253}{{\ttfamily arXiv:hep-ph/0405253
  [hep-ph]}}.

\bibitem{Belanger:2001fz}
G.~Belanger, F.~Boudjema, A.~Pukhov, and A.~Semenov, ``{MicrOMEGAs: A Program
  for calculating the relic density in the MSSM},''
  \href{http://dx.doi.org/10.1016/S0010-4655(02)00596-9}{{\em
  Comput.Phys.Commun.} {\bfseries 149} (2002) 103--120},
\href{http://arxiv.org/abs/hep-ph/0112278}{{\ttfamily arXiv:hep-ph/0112278
  [hep-ph]}}.

\bibitem{2009ApJS..180..330K}
E.~{Komatsu}, J.~{Dunkley}, M.~R. {Nolta}, C.~L. {Bennett}, B.~{Gold},
  G.~{Hinshaw}, N.~{Jarosik}, D.~{Larson}, M.~{Limon}, L.~{Page}, D.~N.
  {Spergel}, M.~{Halpern}, R.~S. {Hill}, A.~{Kogut}, S.~S. {Meyer}, G.~S.
  {Tucker}, J.~L. {Weiland}, E.~{Wollack}, and E.~L. {Wright}, ``{Five-Year
  Wilkinson Microwave Anisotropy Probe Observations: Cosmological
  Interpretation},'' \href{http://dx.doi.org/10.1088/0067-0049/180/2/330}{{\em
  apjs} {\bfseries 180} (Feb., 2009) 330--376},
  \href{http://arxiv.org/abs/0803.0547}{{\ttfamily arXiv:0803.0547}}.

\bibitem{Sekmen:2014vsa}
S.~Sekmen, ``{Inclusive SUSY searches at the LHC},''
\href{http://arxiv.org/abs/1405.4730}{{\ttfamily arXiv:1405.4730 [hep-ex]}}.

\bibitem{Fowlie:2013oua}
{\bfseries BayesFITS Group} Collaboration, A.~Fowlie, K.~Kowalska,
  L.~Roszkowski, E.~M. Sessolo, and Y.-L.~S. Tsai, ``{Dark matter and collider
  signatures of the MSSM},''
  \href{http://dx.doi.org/10.1103/PhysRevD.88.055012}{{\em Phys.Rev.}
  {\bfseries D88} no.~5, (2013) 055012},
\href{http://arxiv.org/abs/1306.1567}{{\ttfamily arXiv:1306.1567 [hep-ph]}}.

\bibitem{Freitas:2014pua}
A.~Freitas, J.~Lykken, S.~Kell, and S.~Westhoff, ``{Testing the Muon g-2
  Anomaly at the LHC},''
\href{http://arxiv.org/abs/1402.7065}{{\ttfamily arXiv:1402.7065 [hep-ph]}}.

\bibitem{Chakraborti:2014gea}
M.~Chakraborti, U.~Chattopadhyay, A.~Choudhury, A.~Datta, and S.~Poddar, ``{The
  Electroweak Sector of the pMSSM in the Light of LHC - 8 TeV and Other
  Data},''
\href{http://arxiv.org/abs/1404.4841}{{\ttfamily arXiv:1404.4841 [hep-ph]}}.

\bibitem{Sjostrand:2006za}
T.~Sjostrand, S.~Mrenna, and P.~Z. Skands, ``{PYTHIA 6.4 Physics and Manual},''
  \href{http://dx.doi.org/10.1088/1126-6708/2006/05/026}{{\em JHEP} {\bfseries
  0605} (2006) 026},
\href{http://arxiv.org/abs/hep-ph/0603175}{{\ttfamily arXiv:hep-ph/0603175
  [hep-ph]}}.

\bibitem{Pumplin:2002vw}
J.~Pumplin, D.~Stump, J.~Huston, H.~Lai, P.~M. Nadolsky, {\em et~al.}, ``{New
  generation of parton distributions with uncertainties from global QCD
  analysis},'' \href{http://dx.doi.org/10.1088/1126-6708/2002/07/012}{{\em
  JHEP} {\bfseries 0207} (2002) 012},
\href{http://arxiv.org/abs/hep-ph/0201195}{{\ttfamily arXiv:hep-ph/0201195
  [hep-ph]}}.

\bibitem{Cacciari:2011ma}
M.~Cacciari, G.~P. Salam, and G.~Soyez, ``{FastJet User Manual},''
  \href{http://dx.doi.org/10.1140/epjc/s10052-012-1896-2}{{\em Eur.Phys.J.}
  {\bfseries C72} (2012) 1896},
\href{http://arxiv.org/abs/1111.6097}{{\ttfamily arXiv:1111.6097 [hep-ph]}}.

\bibitem{Cacciari:2008gp}
M.~Cacciari, G.~P. Salam, and G.~Soyez, ``{The Anti-k(t) jet clustering
  algorithm},'' \href{http://dx.doi.org/10.1088/1126-6708/2008/04/063}{{\em
  JHEP} {\bfseries 0804} (2008) 063},
\href{http://arxiv.org/abs/0802.1189}{{\ttfamily arXiv:0802.1189 [hep-ph]}}.

\bibitem{Kidonakis:2011tg}
N.~Kidonakis, ``{Top Quark Theoretical Cross Sections and pT and Rapidity
  Distributions},''
\href{http://arxiv.org/abs/1109.3231}{{\ttfamily arXiv:1109.3231 [hep-ph]}}.

\bibitem{Campbell:2011bn}
J.~M. Campbell, R.~K. Ellis, and C.~Williams, ``{Vector boson pair production
  at the LHC},'' \href{http://dx.doi.org/10.1007/JHEP07(2011)018}{{\em JHEP}
  {\bfseries 1107} (2011) 018},
\href{http://arxiv.org/abs/1105.0020}{{\ttfamily arXiv:1105.0020 [hep-ph]}}.

\bibitem{Beenakker:1999xh}
W.~Beenakker, M.~Klasen, M.~Kramer, T.~Plehn, M.~Spira, {\em et~al.}, ``{The
  Production of charginos / neutralinos and sleptons at hadron colliders},''
  \href{http://dx.doi.org/10.1103/PhysRevLett.100.029901,
  10.1103/PhysRevLett.83.3780}{{\em Phys.Rev.Lett.} {\bfseries 83} (1999)
  3780--3783},
\href{http://arxiv.org/abs/hep-ph/9906298}{{\ttfamily arXiv:hep-ph/9906298
  [hep-ph]}}.

\bibitem{Aad:2014vma}
{\bfseries ATLAS Collaboration} Collaboration, G.~Aad {\em et~al.}, ``{Search
  for direct production of charginos, neutralinos and sleptons in final states
  with two leptons and missing transverse momentum in pp collisions at
  $\sqrt{s}$ = 8 TeV with the ATLAS detector},''
\href{http://arxiv.org/abs/1403.5294}{{\ttfamily arXiv:1403.5294 [hep-ex]}}.

\bibitem{Khachatryan:2014qwa}
{\bfseries CMS Collaboration} Collaboration, V.~Khachatryan {\em et~al.},
  ``{Searches for electroweak production of charginos, neutralinos, and
  sleptons decaying to leptons and W, Z, and Higgs bosons in pp collisions at 8
  TeV},''
\href{http://arxiv.org/abs/1405.7570}{{\ttfamily arXiv:1405.7570 [hep-ex]}}.

\end{thebibliography}
\providecommand{\href}[2]{#2}\begingroup\raggedright\endgroup
\end{document}